\documentstyle[12pt,a4wide]{article}
\newcommand{\beq}[1]{\begin{equation}\label{#1}}
\newcommand{\eeq}{\end{equation}}
\newcommand{\bear}[1]{\begin{eqnarray}\label{#1}}
\newcommand{\ear}{\end{eqnarray}}
\newcommand{\nn}{\nonumber}

\newcommand{\nl}{ {\hfill \break} }
\newcommand{\np}{ {\newpage } }

\newcommand{\lto}{ {\longrightarrow } }

\def\C{\mbox{\rm {I\kern-.520em C}}}

\newcommand{\diag}{ \mbox{\rm diag} }

\newcommand{\df}{\mbox{\rm d}}

\newcommand{\absdet}[1]{\vert\det{#1}\vert}
\newcommand{\partlx}[1]{\frac{\partial}{\partial x^{#1}}}
\begin{document}
%
\vspace*{-2.0cm}
\centerline{\mbox{\hspace*{13cm} IPM-96-149}}
\centerline{\mbox{\hspace*{13cm} FUB-HEP/96-3}}
\centerline{\mbox{\hspace*{13cm}June 1996}}
\centerline{\mbox{\hspace*{13cm}to appear in:}}
\centerline{\mbox{\hspace*{13cm}Phys. Rev. D}}
\vspace*{2.0cm}
\centerline{\large \bf Tensor-multi-scalar theories from} 
\centerline{\large \bf multidimensional cosmology\footnote{This work
was financially supported by 
DFG grants 436 RUS 113/7/0, 436 UKR 17/7/95,     
and the DAAD.           }     }
\vspace{1.03truecm}

\bigskip

\centerline{\bf \large M. Rainer\S\dag, A. Zhuk{\S\ddag}}

\vspace{0.96truecm}

\centerline{\S Gravitationsprojekt, Institut f\"ur Mathematik}
\centerline{ Universit\"at Potsdam, PF 601553}
\centerline{D-14415 Potsdam, Germany}

\vspace{0.15truecm}

\centerline{\dag Institute for Studies in Physics and Mathematics}
\centerline{Farmanieh Bld., P.O.Box  5531}
\centerline{Tehran 19395, Iran}

\vspace{0.15truecm}

\centerline{\ddag Department of Physics}
\centerline{University of Odessa, 2 Petra Velikogo St.}
\centerline{Odessa 270100, Ukraine}
\vspace{2.0truecm}

\begin{abstract}
Inhomogeneous multidimensional cosmological models
with a higher dimensional  space-time manifold 
$M=M_0\times\prod_{i=1}^n M_i$ ($n\geq 1$) are investigated
under dimensional reduction to tensor-multi-scalar theories.
In the Einstein conformal frame, these theories
take the shape of a flat $\sigma$-model.
For the singular case where $M_0$ is $2$-dimensional, 
the dimensional reduction to dilaton gravity is preformed
with different distinguished representations of the action.  
\end{abstract}
\medskip
\medskip
\medskip
\vspace{1.35truecm}
\hspace*{0.950cm} PACS number(s):\  04.50.+h,\ 98.80.Hw,\ 04.60.Kz
\np

\section{\bf Introduction}
\setcounter{equation}{0}
In recent years, scalar-tensor theories have received a renewed interest.
There are two reasons for it. First, extended inflation \cite{LS1,LS2}
which originally was based on standard Brans-Dicke (BD) theory \cite{BD}
revives the scenario of inflation via a first-order transition.
It provides a natural (non-fine-tuned) way to restore the original ideas
of inflation while avoiding the cosmological difficulties
coming from vacuum-dominated exponential expansion
obtained in general relativity.
Second, scalar tensor theories, generalizing standard BD theory, 
can satisfy the solar system criteria \cite{DT}
and other present observations \cite{DGG}  
to arbitrary accuracy, but still diverge from general relativity
in the strong field limit.
Thus, in future, such theories may provide an important test of general
relativity. It should be noted in that context that, 
via conformal transformation of the metric, we can write scalar-tensor
models equally well in the Jordan-Brans-Dicke frame or in the 
Einstein-Pauli frame, and the question which frame is a physical one
is still open.

Several ways to generalize standard BD theory were proposed.
These theories can be split  into three main groups. First, there are
the theories with BD parameter $\omega$ depending
on the dilatonic scalar field \cite{Wag}-\cite{SA}. The second class
is represented by theories with more than one dilaton field
\cite{DT,DEF,BH}. All other possible variations of the standard
BD model form the third group,
containing e.g. models in which the dilaton couples with different
strengths to both, visible matter and conjectured "dark" matter
\cite{DGG}.

Scalar-tensor theories follow naturally as the low-energy
limit of various Kaluza-Klein theories.
Among them, multidimensional cosmological models with a space-time
consisting of $n\geq 2$  Einstein spaces are most popular.
Usually, theories with one internal Einstein space ($n=2$)
are considered \cite{AKLO}-\cite{MS}. The dimensional reduction of
these theories yields only one dilaton field 
(like in the original BD theory).
Here we show that this case is exceptional, with
a midisuperspace metric of degenerate signature.
The model with $n\geq 2$ was considered in the paper \cite{BDLLU},
where the emphasis was on the problem of 
the internal dimensions  compactification.
In our paper we shall give a more elegant way
of dimensional reduction to tensor-multi-scalar theories
which reveals explicitly the nature of the dilaton fields.
For example, the dilaton field with opposite sign in the kinetic
term of the Lagrangian is connected with a dynamical
volume of the whole internal space.
In a conformal Einstein Pauli frame,
a $\sigma$-model representation
of the theory under consideration can easily be obtained.
   
In Sec. 2, it is shown, 
for space-time dimension $D_0>2$ after dimensional reduction,
that, in the $\sigma$-model representation of our model, 
the metric on the space of scalar fields  
is the flat Euclidean one.
If $D_0=2$, there is no conformal Einstein (Pauli) frame.
This is actually no fault of the theory here, because in two dimensions
the Einstein action is a topological invariant,
whence it does not yield a dynamics of the $2$-geometry.

Nevertheless it is worthwhile to consider here 
$2$-dimensional dilation gravity, which is a subject of intensive
investigations recently \cite{AAR,AADZ,KlStr}.
In Sec. 3 we obtain the action of $2$-dimensional 
dilaton gravity under dimensional reduction from 
cosmological models. Different representations are given, which correspond
to different choices of conformal frames.
\section{\bf Effective BD Models from Multidimensional Models}
\setcounter{equation}{0}
Let us now consider a multidimensional space-time manifold
\beq{0}
M=M_{0}\times\prod_{i=1}^{n} M_{i} .
\eeq
The metric on $M$ can be decomposed as
\beq{1}
g=g^{(0)}+\sum_{i=1}^{n} e^{2\beta^i(x)}g^{(i)} ,
\eeq
where $x$ are some coordinates of $M_0$, and
\beq{1a}
g^{(0)}=g^{(0)}_{\mu\nu}(x) \df x^{\mu} \otimes \df x^{\nu} .
\eeq
With the Laplace-Beltrami operator on $M_0$ defined by
\beq{Lapl}
\Delta[g^{(0)}]=
\frac{1}{\sqrt{\absdet{g^{(0)}}}} 
{\partlx{\mu}}
\left( \sqrt{\absdet{g^{(0)}}} g^{(0)\mu\nu} 
{\partlx{\nu}}
\right) ,
\eeq
we get the Ricci curvature scalar \cite{IM}
\bear{R}
\lefteqn{ R[g]=R[g^{(0)}]+\sum_{i=1}^{n} e^{-2\beta^i} R[g^{(i)}] }
\nn\\
& & - \sum_{i,j=1}^{n} (D_i\delta_{ij}+D_i D_j)  
g^{(0)\mu\nu} 
\frac{\partial\beta^i}{\partial x^\mu} 
\frac{\partial\beta^j}{\partial x^\nu} 
-2 \sum_{i=1}^{n} D_i \Delta[g^{(0)}] \beta^i . 
\ear
With total dimension $D:=\sum_{i=0}^{n} D_i$,
$\kappa^2$ a $D$-dimensional gravitational coupling constant,
and $S_{GH}$ the standard Gibbons-Hawking boundary term,
we consider an action of the form
\beq{Sm}
S=\frac{1}{2\kappa^2}\int_{M} \df^Dx \sqrt{\absdet{g}} R[g]+ S_{GH},
\eeq
In the following we assume that $R[g^{(i)}]$ is finite on
$(M_i,g^{(i)})$.
Let us now consider the volumes $\mu_i$ of $(M_i,g^{(i)})$ and the
total internal space volume $\mu$, satisfying 
\beq{mui}
\mu_i=\int_{M_i}\df^{D_i}y \sqrt{ \absdet{ g^{(i)} } }\ , \quad
\mu=\prod_{i=1}^n \mu_i\ . 
\eeq
If all of  the spaces $M_i$ ($i=1,\ldots,n$) are compact,
then the volumes $\mu_i$ and $\mu$ are finite,
and so are also the numbers
\beq{rhoi}
\rho_i=\int_{M_i}\df^{D_i}y \sqrt{ \absdet{ g^{(i)} } } R[g^{(i)}]  \  . 
\eeq
However, a non-compact $M_i$ might have infinite volume $\mu_i$  or infinite
$\rho_i$.
Nevertheless, for the following we have to
assume only that all ratios
$\frac{\rho_i}{\mu_i}$, $i=1,\ldots,n$ are finite. 
This is in particular the case, when $M_i$
is {\em homogeneous}. Then 
$$
\frac{\rho_i}{\mu_i}=R[g^{(i)}]
$$ 
is always constant and finite.
In the special case, where $M_i$ is a {\em Einstein} manifold
$R^k_{j}[g^{(i)}]=\lambda^i \delta^k_{j}$ with constant $\lambda^i$, it is 
$$
\frac{\rho_i}{\mu_i}=R[g^{(i)}]=\lambda^i D_i,
$$ 
and, more specially, when $M_i$ has {\em constant curvature} $k$,
$$
\frac{\rho_i}{\mu_i}=R[g^{(i)}]=k D_i (D_i -1)\ .
$$
However, here we do not restrict apriori to Einstein or constant curvature
spaces. For convenience and beauty, in the following we will exemplify the 
dimensional reduction just for the case of homogeneous spaces 
$M_1,\ldots,M_n$, although the proceedure could be easily generalized 
for the case of inhomogeneous $M_1,\ldots,M_n$. (Then one 
has to work with $\frac{\rho_i}{\mu_i}$ in place of $R[g^{(i)}]$.)  
The bare gravitational coupling constant  
\beq{k2}
\kappa^2={\kappa_0}^2\cdot \mu
\eeq
might become infinite, while the true $D_0$-dimensional coupling constant
${\kappa_0}^2$ is always finite.
If $D_0=4$, then ${\kappa_0}^2=8\pi G_N$, where $G_N$ is the Newton constant. 
Then the action (\ref{Sm}) remains 
well defined, even when some of the volumes $\mu_i$ are infinite. 
After rewriting 
\bear{part}
\lefteqn{ 
\frac{1}{\kappa^2}
\int_{M} \df^Dx \sqrt{\absdet{g}} \sum_{i=1}^{n} D_i \Delta[g^{(0)}] \beta^i
}\nn\\
&&
=\frac{\mu}{\kappa^2}
\sum_{i=1}^{n} D_i \int_{M_0}\df^{D_0}x \sqrt{\absdet{g^{(0)}}}
\prod_{l=1}^n e^{D_l\beta^l}
\frac{1}{\sqrt{\absdet{g^{(0)}}}} 
\partlx{\lambda}
\left( \sqrt{\absdet{g^{(0)}}} g^{(0)\lambda\nu} 
\partlx{\nu}\beta^i
\right)
\nn\\
&&
=\frac{1}{\kappa^2_0}
\sum_{i=1}^{n} D_i \int_{M_0}\df^{D_0}x 
\left[
\partlx{\lambda}
\left( \sqrt{\absdet{g^{(0)}}} g^{(0)\lambda\nu}
\prod_{l=1}^n e^{D_l\beta^l} 
\partlx{\nu}\beta^i 
\right)
\right.
\nn\\
&&
\hspace*{4.5cm}
\left.
-\sqrt{\absdet{g^{(0)}}} g^{(0)\lambda\nu}
\frac{\partial \beta^i}{\partial x^\nu}
\prod_{l=1}^n e^{D_l\beta^l} 
\sum_{j=1}^{n} D_j \frac{\partial \beta^j}{\partial x^\lambda}
\right]
\nn\\
&&
=S_{GH}
-\frac{1}{\kappa^2_0}
\int_{M_0}\df^{D_0}x 
\sqrt{\absdet{g^{(0)}}}
\prod_{l=1}^n e^{D_l\beta^l} 
\sum_{i,j=1}^{n}
D_i D_j
g^{(0)\lambda\nu}
\frac{\partial \beta^i}{\partial x^\lambda}
\frac{\partial \beta^j}{\partial x^\nu} \ ,
\ear
the action is
\beq{S}
S=\frac{1}{2\kappa^2_0}\int_{M_0} \df^{D_0}x \sqrt{\absdet{g^{(0)}}} 
\prod_{l=1}^n e^{D_l\beta^l} 
\left\{
R[g^{(0)}]
-\sum_{i,j=1}^{n}
G_{ij}
g^{(0)\lambda\nu}
\frac{\partial \beta^i}{\partial x^\lambda}
\frac{\partial \beta^j}{\partial x^\nu}
+\sum_{i=1}^{n}
R[g^{(i)}] e^{-2\beta^i}
\right\},
\eeq
where 
\beq{Gij}
G_{ij}:=D_i \delta_{ij} - D_i D_j.
\eeq
Let us first consider the exceptional case $n=1$.
\beq{S1}
S=\frac{1}{2\kappa^2_0}\int_{M_0} \df^{D_0}x \sqrt{\absdet{g^{(0)}}} 
e^{D_1\beta^1} 
\left\{
R[g^{(0)}]
+D_1 (D_1- 1)
g^{(0)\lambda\nu}
\frac{\partial \beta^1}{\partial x^\lambda}
\frac{\partial \beta^1}{\partial x^\nu}
+R[g^{(1)}] e^{-2\beta^1}
\right\}.
\eeq
Here, for $D_1>1$ the kinetic term  has
a different sign than usual, and for $D_1=1$
there is no kinetic term at all.
Setting 
\beq{phi1}
\phi:=e^{D_1\beta^1},
\eeq
it is 
$\frac{\partial \beta^1}{\partial x^\lambda}=
\frac{1}{D_1}\frac{1}{\phi}\frac{\partial \phi}{\partial x^\lambda}$,
and hence
\beq{S1BD}
S=\frac{1}{2\kappa^2_0}\int_{M_0} \df^{D_0}x \sqrt{\absdet{g^{(0)}}} 
\left\{
\phi R[g^{(0)}]
-\omega
g^{(0)\lambda\nu}
\frac{1}{\phi}
\frac{\partial \phi}{\partial x^\lambda}
\frac{\partial \phi}{\partial x^\nu}
+R[g^{(1)}]  \phi^{1-\frac{2}{D_1}}
\right\},
\eeq
with BD parameter $\omega=\omega(D_1)=(\frac{1}{D_1}-1)$,
depending on the present extra dimension $D_1$.
It is remarkable that, the conformal
coupling constant ${\xi_{c,d+1}}$ in dimension $d+1$ 
determines the BD parameter for general extra dimension $d$ as   
\beq{omega}
\omega(d):=\frac{1}{d}-1\equiv-{4} {\xi_{c,d+1}}\ .
\eeq
Let us now examine the general case $n>1$.
Here it is useful to diagonalize the metric tensor (\ref{Gij}).
For the midisuperspace metric 
\bear{G}
G:=G_{ij}\df\beta^i\otimes \df\beta^j
=\eta_{kl}\df z^k\otimes \df z^l=-\df z^1\otimes \df z^1+\sum_{i=2}^{n}
\df z^i\otimes \df z^i\ ,
\ear
the diagonalizing transformation 
\beq{diag}
z^i=T^i{}_j \beta^j\ ,\quad i=1,\ldots,n 
\eeq
is  given by (see also \cite{IMZ}) 
\bear{z}
z^1&=&q^{-1}\sum_{j=1}^{n}D_j\beta^j\ ,
\nn\\
z^i&=&{\left[\left.D_{i-1}\right/\Sigma_{i-1}\Sigma_{i}\right]}^{1/2}
\sum_{j=i}^{n}
D_j\left(\beta^j-\beta^{i-1}\right)\ ,
\ear
$i=2,\ldots,n$, where
$$
q:={\left[\left.{D'}\right / ({D'}-1)\right]}^{1/2}=
\frac{1}{2{\sqrt{\xi_{c,D'+1}}}}\ ,\quad
{D'}:=D-D_0\ ,\quad
\Sigma_k:=\sum_{i=k}^{n}D_i\ .
$$
Especially, we have
\beq{T1}
T^1{}_i=\frac{D_i}{q}\ ,\quad i=1,\ldots,n
\eeq
Let us determine $U=T^{-1}$ inverting Eq. (\ref{diag}) to
\beq{undiag}
\beta^i=U^i{}_j z^j\ ,\quad i,j=1,\ldots,n\ .
\eeq
Eqns. (\ref{G}) and  (\ref{diag}) imply
$G_{ij}=\eta_{kl}T^k{}_i T^l{}_j$, $i,j=1,\ldots,n$,
and hence
\beq{Tinv}
U^i{}_j=G^{ik} T^l{}_k \eta_{lj}=G^{ik} (T^t)_k{}^l \eta_{lj},\quad 
i,j=1,\ldots,n,
\eeq
where the tensor components of the inverse midisuperspace
metric are given as
\beq{invG}
G^{ij}=\frac{\delta^{ij}}{D_i}+\frac{1}{1-D'}.
\eeq
With (\ref{T1}), we obtain especially 
\beq{invT}
U^i{}_1=G^{ij} T^k{}_j \eta_{k1}=-G^{ij} T^1{}_j=\frac{1}{q(D'-1)}
=\frac{1}{\sqrt{D'(D'-1)}},\quad 
i=1,\ldots,n.
\eeq
Using that, we can rewrite the action 
(\ref{S}) as
\bear{Sz}
S&=&\frac{1}{2\kappa^2_0}\int_{M_0} \df^{D_0}x \sqrt{\absdet{g^{(0)}}} 
\prod_{l=1}^n e^{D_l\beta^l} 
\left\{
R[g^{(0)}]
+g^{(0)\lambda\nu}
\frac{\partial z^1}{\partial x^\lambda}
\frac{\partial z^1}{\partial x^\nu}
-\sum_{i=2}^{n}
g^{(0)\lambda\nu}
\frac{\partial z^i}{\partial x^\lambda}
\frac{\partial z^i}{\partial x^\nu}
\right.
\nn\\
&&
\hspace*{6.50cm}
\left.
+(e^{q z^1})^{-\frac{2}{D'}}
\sum_{i=1}^{n}
R[g^{(i)}] e^{-2\sum_{k=2}^{n}U^i_k z^k}
\right\}.
\ear
Let define the BD field as
\beq{BD}
\phi:=e^{q z^1}
=\prod_{l=1}^n e^{D_l\beta^l}
\equiv v_{\mbox{\rm int}}\ ,
\eeq
where
\beq{vint}
v_{\mbox{\rm int}}:=V_{\mbox{\rm int}}/\mu 
\eeq
is a scale which renormalizes the internal space volume  
$V_{\mbox{\rm int}}:=
\int_{M_1\times\cdots\times M_n}\df^{D'}y \sqrt{ \absdet{(g-g^{(0))} } } $.
Its corresponding logarithmic scale factor is the dilaton field $z^1$.
The  derivative of the latter is
\beq{dz1}
\partlx{\mu} z^1
=\frac{1}{q\phi} \partial_{\mu}\phi\ .
\eeq
So we can write the action (\ref{Sz}) as
\bear{SPhi}
S&=&\frac{1}{2\kappa^2_0}\int_{M_0} \df^{D_0}x \sqrt{\absdet{g^{(0)}}} 
\left\{
\phi R[g^{(0)}]
-\omega g^{(0)\lambda\nu}
\frac{1}{\phi}
\frac{\partial \phi}{\partial x^\lambda}
\frac{\partial \phi}{\partial x^\nu}
-\phi\sum_{i=2}^{n}
g^{(0)\lambda\nu}
\frac{\partial z^i}{\partial x^\lambda}
\frac{\partial z^i}{\partial x^\nu}
\right.
\nn\\
&&
\hspace*{5.0cm}
\left.
+\phi^{1-\frac{2}{D'}}
\sum_{i=1}^{n}
R[g^{(i)}] e^{-2\sum_{k=2}^{n} U^i_k z^k}
\right\}\ ,
\ear
where now $\omega=\omega(D')=\frac{1}{D'}-1\equiv-4\xi_{c,D'+1}$
is the BD parameter, depending now on the total extra dimension $D'$.
In the action (\ref{S}) all scalar fields $\beta^i$, $i=1,\ldots,n$ 
couple to the curvature $R[g^{(0)}]$. After the diagonalization 
(\ref{diag}) only one of the scalar fields, namely the BD field $\phi$,
is coupled to the curvature. In the action (\ref{SPhi})
scalar fields $z^i$ play the role of normal scalar matter fields
coupled to the dilaton BD scalar $\phi$.
Note that the kinetic terms for the fields $z^i$, $i=2,\ldots,n$,
have the usual normal sign.
In contrast to the action (\ref{S1BD}) w.r.t. its field $\phi$,
Eqn. (\ref{SPhi}) contains {\em no} selfinteraction terms for any of its 
fields  $\phi$ and $z^i$, $i=2,\ldots,n$.
Rather it contains $\phi-z^i$ cross terms ($i=2,\ldots,n$) !
These cross terms are, like the fields $z^i$ and $\phi$ themselves, 
of purely geometric nature.             
The exceptional case (\ref{S1BD}) corresponds formally to the case
$z^k\equiv 0,\ D_k\equiv 0 $ ($k=2,\ldots,n$) of (\ref{SPhi}).

For $D_0\neq 2$, the action (\ref{SPhi}) can be written in a $\sigma$-model
representation \cite{DEF}. We define a new metric $\hat g^{(0)}_{\mu\nu}$,
which yields the so called {\em Einstein conformal frame},
and new scalar fields $\varphi^i$ ($i=1,\ldots,n$) by
\bear{sigmaframe}
\hat g^{(0)}_{\mu\nu}&=&\phi^{\frac{2}{D_0-2}} g^{(0)}_{\mu\nu}\ ,
\nn\\
\varphi^1&=&-A\ln\phi\ ,
\nn\\
\varphi^i&=&z^i,\quad i=2,\ldots,n\ , 
\ear
where $A:=\pm\left[
\omega(D')+\frac{D_0-1}{D_0-2}
\right]^{\frac{1}{2}}$.
Note that, this transformation is regular for  
$\omega(D')\neq \omega_{c,D_0}$, where
$\omega_{c,D_0}:=-\frac{D_0-1}{D_0-2}\equiv-\frac{1}{4}\xi^{-1}_{c,D_0}$
is the conformal parameter for dimension $D_0$.
Taking into account that $-1<\omega(D')\leq 0$ for $D'\geq 1$ and 
$\omega(0)=\infty$, one obtains: 
If $D_0>2$, (\ref{sigmaframe}) is regular for any $D'> 0$, with $A^2>0$.
For $D_0=2$ or  $D'=0$ (\ref{sigmaframe}) is singular. It is  singular
with $A^2=0$ if $(D_0,D')=(0,2)$.
If $D_0=1$, (\ref{sigmaframe}) is singular for $D'=1$, and
regular for any $D'>1$.
In the latter case $A^2<0$, and a real redefinition of the complex field,
e.g. $\varphi^1\to\vert\varphi^1\vert$, yields again
a Minkowskian metric in the space of scalar fields.  

For $D_0>2$, with the flat $\sigma$-model metric
\bear{sigmag}
\df\sigma=\sigma_{ij}\df\varphi^i\otimes\df\varphi^j,\quad 
(\sigma_{ij})=\diag(+1,\ldots,+1), 
\ear
where $i,j=1,\ldots,n$,
and the potential
\bear{sigmaV}
V(\varphi^i)=-e^{-\frac{B}{A} \varphi^1} \sum_{i=1}^{n} R[g^{(i)}] 
e^{-2\sum_{k=2}^{n} U^i_k\varphi^k}\ ,
\ear
where $B:=1-\frac{2}{D'}-\frac{D_0}{D_0-2}$,
the action (\ref{SPhi}) then reads 
\bear{Ssigma}
S=\frac{1}{2\kappa^2_0}\int_{M_0} \df^{D_0}x \sqrt{\absdet{\hat g^{(0)}}} 
\left\{
\hat R[\hat g^{(0)}]
-\sum_{i,j=1}^{n}\sigma_{ij}\hat g^{(0)\lambda\nu}
\frac{\partial \varphi^i}{\partial x^\lambda}
\frac{\partial \varphi^j}{\partial x^\nu}
-V(\varphi^i)
\right\}.
\ear
Note that the $\sigma$-model metric (\ref{sigmag}) is flat like
the midisuperspace metric (\ref{G}), however, while  (\ref{G}) is
Minkowskian,  (\ref{sigmag}) is Euclidean. So we found 
equivalent representations (\ref{S}) and (\ref{Ssigma}) of the same 
action $S$, but with different signature in their respective space 
of scalar fields.

In the case $n=1$, with just one dilaton $\varphi$,  
the action (\ref{Ssigma}) is equal to  
\bear{Ssigma1}
S=\frac{1}{2\kappa^2_0}\int_{M_0} \df^{D_0}x \sqrt{\absdet{\hat g^{(0)}}} 
\left\{
\hat R[\hat g^{(0)}]
-\hat g^{(0)\mu\nu}
\frac{\partial \varphi}{\partial x^\mu}
\frac{\partial \varphi}{\partial x^\nu}
+ R[g^{(1)}] e^{-\frac{B}{A} \varphi} 
\right\}.
\ear
This action can be written in the 
'string-like' form 
(see e.g. \cite{GSW,MSh,IM2} and refs. therein)
\bear{Sstring1s}
S=\frac{1}{2\kappa^2_0}\int_{M_0} \df^{D_0}x \sqrt{\absdet{\hat g^{(0)}}} 
\left\{
\hat R[
\hat g^{(0)}]
-\hat g^{(0)\mu\nu}
\frac{\partial \varphi}{\partial x^\mu}
\frac{\partial \varphi}{\partial x^\nu}
-2\Lambda e^{-2\lambda \varphi} 
\right\},
\ear
where the constants are fixed by
the conditions
\bear{constants}
2\Lambda&:=&-R[g^{(1)}]\nn\\
\lambda^2&:=&\lambda^2_c=\frac{D-2}{D_1(D_0-2)}.
\ear
In  equation (\ref{Sstring1s}) $\lambda$ is the dilatonic
coupling constant. 
For $D_0=10$ and $\Lambda=0$  (e.g. for a Ricci flat internal space), 
this action describes the scalar-tensor (i.e. Yang-Mills free) part of 
the bosonic sector from the $10$-dimensional Einstein-Yang-Mills 
supergravity that occurs as low energy limit from superstring theory.

For arbitrary $\Lambda\neq 0$,
the action (\ref{Sstring1s}) 
corresponds to the scalar-tensor sector of an effective string action in
dimension $D_0$, only if the dilatonic coupling is fixed to
\bear{slambda}
\lambda^2:=\lambda^2_s=\frac{1}{D_0-2}.
\ear
The coupling (\ref{slambda}) is obtained for our models with 
(\ref{constants}) only in the limit of infinite internal dimension,
\bear{lambdalim}
\lambda^2_c\lto\lambda^2_s=\frac{1}{D_0-2}\quad\mbox{for}\quad D_1\to\infty.
\ear
Especially for the $10$-dimensional effective action, 
the required value of $\lambda^2=\frac{1}{8}$ is obtained just
in this limit,
while  above for $\Lambda=0$ 
the value of $\lambda$ was completely arbitrary.
Indeed,  $\Lambda=0$ is a critical value for
the string theories, whence $\Lambda\neq 0$ occurs just for
non-critical string theories.
  
The action  (\ref{Sstring1s}) 
can equivalently be obtained from a multidimensional
cosmological model with a usual cosmological term $\Lambda$,
if the internal space $M_1$ is a Ricci flat Einstein space, i.e. $R[g^1]=0$.
Then the equivalence to our previous model is given by exchanging 
$D-1\longleftrightarrow 1-D_1$, which obviously leaves $D_0$ invariant.
In this case $\lambda^2_c=\frac{D_1}{(D-2)(D_0-2)}$,
and the correspondence to non-critical string theories 
is again given in the limit (\ref{lambdalim}).

Finally note that, for $D_0\to 2$, both couplings $\lambda^2_s$ and
$\lambda^2_c$ become asymptotically equal to $(D_0-2)^{-1}$.
Hence, in the limit  $D_0\to 2$ our models become 
independently from the internal dimension $D_1$ equivalent
to effective low energy models of string theory,
for any scalar curvature $-\Lambda/2$ of the internal space.
\section{\bf $2$d dilaton gravity from inhomogeneous cosmology}
\setcounter{equation}{0}
Let us now consider in more details the dimensional reduction 
to a space-time of dimension $D_0=2$. In this case the conformal
transformation (\ref{sigmaframe}) is singular, whence 
the model of (\ref{SPhi}) can not be expressed in
a conformal Einstein Pauli frame. This is not a fault of the theory,
but rather corresponds to the well known fact that $2$-dimensional Einstein
equations are empty, i.e. they do not imply a dynamics \cite{AAR,AADZ}.
Thus we shall consider $2$-dimensional dilaton gravity only.

We start with the case with one dilaton, $n=1$. 
The action (\ref{S1}) can be written in the 
'string-like' form \cite{CGHS,Mi,MiSch} 
\beq{S12}
S=\frac{1}{2\kappa^2_0}\int_{M_0} \df^{2}x \sqrt{\absdet{g^{(0)}}} 
e^{-2\sigma} 
\left\{
R[g^{(0)}]
+4m
g^{(0)\lambda\nu}
\frac{\partial \sigma}{\partial x^\lambda}
\frac{\partial \sigma}{\partial x^\nu}
-2\Lambda e^{-2(\frac{1}{k}+m)\sigma}
\right\},
\eeq
where
\bear{string}
\sigma&:=& -\frac{1}{2}D_1\beta^1\ ,
\nn\\
m&:=& \frac{D_1- 1}{D_1}\ ,
\nn\\
k&:=& -\frac{D_1}{D_1+ 1}\ ,
\nn\\
2\Lambda&:=&-R[g^{(1)}]\ .
\ear
By a conformal transformation of ${g}^{(0)}_{\mu\nu}$ to
\bear{2dsigmaconf}
{\tilde g}^{(0)}_{\mu\nu}&=&e^{-2m\sigma} g^{(0)}_{\mu\nu}\ ,
\ear
we can formulate the action without kinetic dilation term, as
\beq{S12dila}
S=\frac{1}{2\kappa^2_0}\int_{M_0} \df^{2}x \sqrt{\absdet{\tilde g^{(0)}}} 
e^{-2\sigma} 
\left\{
{\tilde R}[\tilde g^{(0)}]
-2\Lambda e^{-\frac{2}{k}\sigma}
\right\}.
\eeq
The $2$d actions (\ref{S12}) and (\ref{S12dila}) are 
invariant under homogeneous conformal transformations 
\bear{constconf}
{\check g}^{(0)}_{\mu\nu}&:=&\Omega^{-2} {\tilde g}^{(0)}_{\mu\nu}\ ,
\nn\\
{\check g}^{(1)}_{\mu\nu}&:=&\Omega^{-2} {g}^{(1)}_{\mu\nu}\ ,
\ear
where $\Omega$ is constant.
Applying (\ref{constconf}) with 
$$
\Omega^2:=-\frac{D_1}{(D_1+1)^{1+\frac{1}{D_1}}} {\frac{1}{2\Lambda}}
$$
yields  
\beq{R1const}
{2\check\Lambda}:=-{\check R}[{\check g}^{(1)}]
=-\frac{D_1}{(D_1+1)^{1+\frac{1}{D_1}}}=\frac{k}{(k+1)^{1+\frac{1}{k}}}
\eeq
and the action (\ref{S12dila}) now reads
\beq{S12fixdila}
S=\frac{1}{2\kappa^2_0}\int_{M_0} \df^{2}x \sqrt{\absdet{\check g^{(0)}}} 
e^{-2\sigma} 
\left\{
{\check R}[\check g^{(0)}]
-{2\check \Lambda} e^{-\frac{2}{k}\sigma}
\right\}.
\eeq
If we assume that the dilaton field is specifically given through the 
geometry on $M_0$ and the dimension $D_1$ of $M_1$, according to
\beq{dilafix}
e^{-2\sigma}:=(k+1) \left( {\check R}[\check g^{(0)}] \right)^k,
\eeq
then the action (\ref{S12fixdila}) takes the form 
\cite{KlStr,Mi,MiSch,MaFFra}
\bear{SMaFFia}
S&=&\frac{1}{2\kappa^2_0}\int_{M_0} \df^{2}x \sqrt{\absdet{\check g^{(0)}}} 
\left( {\check R}[\check g^{(0)}] \right)^{k+1}
\nn\\
&=&\frac{1}{2\kappa^2_0}\int_{M_0} \df^{2}x \sqrt{\absdet{\check g^{(0)}}} 
\left( {\check R}[\check g^{(0)}] \right)^{\frac{1}{D_1+1}}.
\ear
In the general case of multi-scalar fields, the kinetic term of the dilaton
can be removed by an obvious analogous proceedure.
The 'string-like' form of the action (\ref{Sz}) is  
\bear{Szstring}
S&=&\frac{1}{2\kappa^2_0}\int_{M_0} \df^{2}x \sqrt{\absdet{g^{(0)}}} 
e^{-2\sigma} 
\left\{
R[g^{(0)}]
+4m g^{(0)\lambda\nu}
\frac{\partial \sigma}{\partial x^\lambda}
\frac{\partial \sigma}{\partial x^\nu}
-\sum_{i=2}^{n}
g^{(0)\lambda\nu}
\frac{\partial z^i}{\partial x^\lambda}
\frac{\partial z^i}{\partial x^\nu}
\right.
\nn\\
&&
\hspace*{6.50cm}
\left.
-e^{-2(\frac{1}{k}+m)\sigma}
\sum_{i=1}^{n}
2\Lambda_i e^{-2\sum_{j=2}^{n}U^i_j z^j}
\right\},
\ear
where now
\bear{stringmulti}
\sigma&:=& -\frac{1}{2}q z^1\ ,
\nn\\
m&:=& \frac{1}{q^2}=\frac{D'- 1}{D'}\ ,
\nn\\
k&:=& -\frac{D'}{D'+ 1}\ ,
\nn\\
2\Lambda_i&:=&-R[g^{(i)}].
\ear
With (\ref{stringmulti}), the conformal transformation (\ref{2dsigmaconf})  
yields
\bear{Szstringdila}
S&=&\frac{1}{2\kappa^2_0}\int_{M_0} \df^{2}x \sqrt{\absdet{\tilde g^{(0)}}} 
e^{-2\sigma} 
\left\{
\tilde R[\tilde g^{(0)}]
-\sum_{i=2}^{n}
\tilde g^{(0)\lambda\nu}
\frac{\partial z^i}{\partial x^\lambda}
\frac{\partial z^i}{\partial x^\nu}
\right.
\nn\\
&&
\hspace*{6.50cm}
\left.
-e^{-\frac{2}{k}\sigma}
\sum_{i=1}^{n}
2\Lambda_i e^{-2\sum_{j=2}^{n}U^i_j z^j}
\right\}.
\ear
In (\ref{Szstringdila}) there is no kinetic term of the dilaton field.
The kinetic terms of all extra scalar fields $z^i$ have the normal  
sign. The extra fields $z^i$ play the role of usual matter, 
coupling to the dilaton field $\sigma$. 
\section{\bf Conclusions and Discussion}
\setcounter{equation}{0}
We started from multidimensional cosmology. The corresponding 
metric is, from one side, a generalization of the Friedmann metric, which
corresponds here to the special case where all $M_0,M_1,\ldots,M_n$
are spaces of constant curvature. From another side, our 
metric generalizes the anisotropic Kasner metric. In contrast to the
(spatially) homogeneous Friedmann and Kasner metrics, our multidimensional 
metric is in general a (spatially) inhomogeneous one with scale factors
depending on spatial coordinates of $M_0$. 
We obtained effective BD formulations for multidimensional models
via dimensional reduction on $M_0$.
Selfinteraction terms appear exclusively in the degenerate case (\ref{S1BD}) 
where there is only one scalar field.
For $n\geq 2$ scalar fields, the BD like effective action (\ref{SPhi}) 
contains $\phi-z^i$
cross terms, between the BD field $\phi$ and the other scalar fields
$z^i$, $i=2,\ldots,n$, instead.

In the case of only one internal space, $M_1$, the actions obtained after
dimensional reduction of the multidimensional Einstein-Hilbert one
may be written in string-like form. Thus, the associated field
equations have the same form as for the (scalar) bosonic sector 
of the superstring theory in the low energy limit.  
The corresponding effective models of string theory are obtained from our
models in the limit of infinite internal dimension $D_1\to\infty$.

The BD like effective action (\ref{SPhi}), which has a Minkowskian metric
in the space of  scalar fields, is (with few exceptional cases)
equivalent to a conformal Einstein $\sigma$-model action
(\ref{Ssigma}), which has an Euclidean metric in the space of 
scalar fields.
The case of a $1$-dimensional universe
is exceptional: There the metric in the space of (real) scalar fields
of a conformal  Einstein $\sigma$-model is also Minkowskian.

With the effective dimension of the universe $D_0$ and the total
extra dimension $D'$,
the singular cases of the conformal transformation are given by 
$D_0=2$ or $D'=0$, where 
(\ref{sigmaframe}) is undefined, or by 
$D=D_0+D'=2$, where
$A$ in (\ref{sigmaframe}) is zero.
In these exceptional cases our model is not conformal
to an Einsteinian one.  
However, one should also keep in mind that Einstein equations in a 
$2$-dimensional space-time do not imply any dynamics 
(see \cite{AAR,AADZ}). 
For a space-time with $D_0=2$ 
the dimensional reduction of the multidimensional model can be written
as a  'string-like' dilaton gravity, representable in the form 
(\ref{Szstringdila}), where the dilaton appears without kinetic term,
and all extra fields couple to the dilaton with normal signs
of their kinetic terms. If there are no fields besides the dilaton,
then the action can be represented in the form
(\ref{SMaFFia}) (see also \cite{KlStr,Mi,MiSch,MaFFra}),
which has a non-trivial variation only for nonvanishing 
extra dimension $D_1>0$.

A conformal equivalence transformation between two scalar-tensor 
Lagrangian models becoming singular at specific parameters 
(here given by the exceptional dimensions) is a
familiar effect. Such singularities
yielding inequivalent models were also discussed in \cite{Ra}.

Although, in the exceptional dimensions, the models (\ref{SPhi}) and
(\ref{Ssigma}) are mathematically inequivalent, the question
remains, as for all other dimensions, which model is the physical one. 
The difference in the exceptional cases is that, in principle
this question could be decided by experiments on a {\em classical} level. 
For the dimension $D_0>2$ the two
models are mathematically equivalent; so on the classical level
it can not be decided which is the physical one. However,
if one demands that the gravitational interaction is generated
by a pure massless spin-$2$ graviton (without scalar spin-$0$ admixture), 
then, reasoning similar  as in \cite{Cho}, 
(\ref{Ssigma}) rather than (\ref{SPhi}) has to be taken
as the physical model.

Taking into account the conformal relation of scalar-tensor theories to 
fourth (or higher) order gravity (see e.g. \cite{MaSo}), the recent debate 
on the physical metric \cite{Cots1}-\cite{Cots2} concerns also the
corresponding scalar-tensor theories. The result of this purely classical
debate was rather poor: It mainly confirms Brans \cite{Bra},
who pointed out that, once the weak equivalence principle 
holds true in a given frame (in \cite{Bra} it is the frame of 
the original higher order gravity), it will be violated
in any non-trivially conformally related frame. However the choice
of the frame with respect to which a test particle of ordinary matter moves   
along geodesics remains arbitrary for classical scalar-tensor theories.
So the frame of Brans and Dicke \cite{BD} might be the physical one,
giving geodesic paths for minimally coupled test matter, or likewise the
Einstein Pauli frame might be the physical one.
In \cite{Haw} Hawking argued that, black holes might follow
geodesics in the Einstein Pauli frame, but violate the  
{\em strong} equivalence principle in the BD frame, 
while the latter provides geodesic paths for usual test matter. 
For massive objects like black holes, this phenomenon is known as the 
Nordvedt effect \cite{Nor}.
Furthermore Cho \cite{Cho2} showed that in the BD frame
quantum corrections enforce also a violation of the {\em weak} equivalence 
principle.
We believe therefore, that the issue of the physical frame will
be resolved finally only by a quantum theory of gravity. 
Since such a theory might not be subject to any equivalence principle,
the latter might no longer serve as guiding principle for the 
physical metric.
However, generalized arguments of Cho \cite{Cho,Cho2}, give 
some hint that the Einstein Pauli frame (when quantum corrections are small
enough not to destroy any frame at all !)
might then nevertheless be taken as the physical one.

It should however be noted that our multi-scalar-tensor theories differ 
essentially from usual scalar-tensor theories: 
There, some "ordinary" matter field is minimally coupled to the geometry,
either in the Jordan Brans Dicke frame or, equally well,  
in the Einstein Pauli frame (see also  \cite{MaSo}).
We saw above that, arguing on the basis of a classical equivalence principle 
for the ordinary matter only, there is no way select the physical frame.
However, in our models all scalar fields are derived from a multidimensional
geometry, which determines {\em all} couplings of {\em all} scalar fields
to the geometry and among each other. These couplings
can be tested in principle by experiments, thus selecting the
physically admissible multi-scalar-tensor theories and their 
corresponding multidimensional counterpart. 
Because of this predictive power, it is tempting to postulate that any 
multi-scalar-tensor model 
should derive its (scalar) fields from a higher dimensional geometry,
i.e. all (scalar) matter should have some geometric origin.
\nl\nl
{\Large {\bf Acknowledgements}}
\nl\nl
This work was supported 
in part by DAAD (M. R.), by DFG grant 436 UKR - 17/7/93 (A. Z.) 
and DFG grant RUS 113/7/0. M. R.,
appreciating the hospitality of colleagues at IPM in Tehran,
is grateful for discussions
with V. Karimipour,  A. Mustafazade, and especially H. Salehi
for useful comments concerning the principle of equivalence.
A. Z. also thanks Prof. Kleinert and the Freie Universit\"at Berlin,
Prof. H. v. Borzeszkowski and Technische Universit\"at
Berlin,
as well as the members of the Gravitationsprojekt at
Universit\"at Potsdam for their hospitality. We are also grateful
to H.-J. Schmidt for useful comments on the subject.

\end{document}